\UseRawInputEncoding
\documentclass[submission,copyright,creativecommons]{eptcs}
\providecommand{\event}{Post Proceedings of ThEdu 2024} 

\usepackage{graphicx}
\usepackage{iftex}

\ifpdf
  \usepackage{underscore}         
  \usepackage[T1]{fontenc}        
\else
  \usepackage{breakurl}           
\fi

\begin{document}

\title{GNU Aris: a web application for students}
\def\titlerunning{GNU Aris: a web application for students}
\author{
Saksham Attri
\institute{Birla Institute of Technology and Science Pilani, Hyderabad Campus, India}
\email{sakshamattri3@gmail.com}
\and
Zolt\'an Kov\'acs
\institute{Private University of Education, Diocese Linz, Austria}
\email{zoltan.kovacs@ph-linz.at}
\and
Aaron Windischbauer
\institute{Private University of Education, Diocese Linz, Austria}
\email{aaron.windischbauer@ph-linz.at}
}
\def\authorrunning{S.~Attri et al.}
\maketitle              
\begin{abstract}
We report on recent improvements to the free logic education software tool GNU Aris,
including the latest features added during the Google Summer of Code 2023 project.
We focused on making GNU Aris a web application to enable almost all users to use it as a
standalone offline web application written in a combination of HTML, JavaScript, and WebAssembly.
We used the Qt Quick framework with Emscripten to compile the application to WebAssembly.
In the report we summarize the user feedback of university students given during a course on logic.
\end{abstract}

\section{Introduction}

GNU Aris,\footnote{\url{https://www.gnu.org/software/aris/manual/aris.pdf}} according to its official web page at \url{https://www.gnu.org/software/aris}, is a logical proof program that supports propositional and predicate logic, as well as Boolean algebra and arithmetical logic, in the form of abstract sequences. Aris uses standard logical symbols and a natural deduction interface, making it easy for beginners to work with. Aris, a part of GNU's
software library,\footnote{\url{https://www.fsf.org}} is free software in the philosophy of GNU's General Public Licence.

Aris was programmed by Ian Dunn, a GNU activist himself, from 2011 to 2014. He made his latest contributions in September 2017 by updating the build system and cleaning up the project. Aris was further developed mainly by the second author during the following years by maintaining and adding minor improvements and also inviting student contributors at his institution at 
the Private University of Education, Diocese Linz
(PHDL), Austria. Aris has also received valuable contributions over the years from enthusiastic users in the form of patches in its official mailing list.

The program for its main application use is designed to manage a list of statements, numbered from 1, consecutively. The first statements are called \textit{premises}, then (after a horizontal line in the desktop version), the \textit{conclusions} are listed. Conclusions can be drawn by using the premises and eventually some formerly given conclusions, by applying \textit{rules}, which are divided into 5 groups: Inference, Equivalence, Predicate, Boolean, and Miscellaneous:
\begin{itemize}
\item \textit{Inference} rules, that work on basic logic, are
\begin{itemize}
\item Modus Ponens ($P\to Q$ and $P$ implies $Q$),
\item Addition ($P$ implies $P\vee Q\vee R\vee\ldots$),
\item Simplification ($P\wedge Q\wedge R\wedge\ldots$ implies $P$ (or $Q$, or $R$, or $\ldots$)),
\item Conjunction ($P$ and $Q$ and $R$ and $\ldots$ implies $P\wedge Q\wedge R\wedge\ldots$),
\item Hypothetical Syllogism (e.g.~$P\to Q$, $R\to S$ and $Q\to R$ implies $P\to S$),
\item Disjunctive Syllogism (e.g.~$\neg P$, $P\vee Q\vee R$ and $\neg R$ implies $Q$),
\item Excluded Middle ($P\vee \neg P$ always holds), and
\item Constructive Dilemma (e.g.~$P\to R$, $P\vee Q$ and $Q\to S$ implies $R\vee S$),
\end{itemize}
8 rules.
\item \textit{Equivalence} rules, that work on sentence parts, include 
\begin{itemize}
\item Implication ($P\to Q$ is equivalent to $\neg P\vee Q$),
\item DeMorgan (e.g.~$\neg(P\wedge Q)$ is equivalent to $\neg P\vee\neg Q$),
\item Association (e.g.~$P\wedge (Q\wedge R)$ is equivalent to $P\wedge Q\wedge R$),
\item Commutativity (e.g.~$P\wedge Q\wedge R$ is equivalent to $Q\wedge R\wedge P$),
\item Idempotence (e.g.~$P \wedge P \wedge Q \wedge R \wedge R \wedge R$ is equivalent to  $P \wedge Q \wedge R$),
\item Distribution (e.g.~$P\wedge(Q\vee R)$ is equivalent to $(P\wedge Q) \vee (P\wedge R)$),
\item Equivalence ($P\leftrightarrow Q$ is equivalent to $(P\to Q)\wedge(Q\to P)$),
\item Double Negation ($\neg\neg P$ is equivalent to $P$),
\item Exportation ($(P\wedge Q)\to R$ is equivalent to $P\to (Q\to R)$), and 
\item Subsumption ($P \wedge (P \vee Q)$ is equivalent to $P$),
\end{itemize}
10 rules.
\item \textit{Predicate} rules, that work with predicates and quantifiers, are
\begin{itemize}
\item Universal Generalization (if $a$ is arbitrary then $P(a)$ implies $\forall x(P(x))$),
\item Universal Instantiation ($\forall x(P(x))$ implies $P(a)$),
\item Existential Generalization ($P(a)$ implies $\exists x(P(x))$,
\item Existential Instantiation (if $a$ is not yet used, $\exists x(P(x))$ implies $P(a)$),
\item Bound Variable (e.g.~$\forall x(P(x))$ is equivalent to $\forall y(P(y))$),
\item Null Quantifier ($\forall x(P(a))$ is equivalent to $P(a)$),
\item Prenex (e.g.~$\exists x(P(x) \wedge Q(a))$ is equivalent to $\exists x(P(x)) \wedge Q(a)$),
\item Identity ($a=a$ always holds), and
\item Free Variable ($a=b$ and $P(a)$ implies $P(b)$),
\end{itemize}
9 rules.
\item \textit{Boolean} rules, that handle
Boolean Algebra, include 
\begin{itemize}
\item Boolean Identity (e.g.~$A\wedge\top$ is equivalent to $A$), 
\item Boolean Negation (e.g.~$A\wedge\neg A$ is equivalent to $\bot$), 
\item Boolean Dominance (e.g.~$A\wedge\bot$ is equivalent to $\bot$), and 
\item Symbol Negation (e.g.~$\neg\top$ is equivalent to $\bot$),
\end{itemize}
4 rules.
\item \textit{Miscellaneous} rules, that do not fit in the above categories, are
\begin{itemize}
\item Lemma,
\item Subproof,
\item Sequence, and 
\item Induction,
\end{itemize}
4 rules.
\end{itemize}
A detailed explanation of these 35 rules can be found in the manual of GNU Aris.

In general, the user aims at finding a derivation of a given conclusion if a set of premises are also given, by entering a sequence of conclusions based on the premises and the previous conclusions by applying the allowed rules. Usually, just a small set of these rules is required to find a derivation for a given example. But the available set of rules is rich enough to build acceptably long derivations with some freedom of the chosen rules. Therefore, it seems that GNU Aris is a good choice to help students to do experiments with building their own proofs, which can be checked by the program syntactically and logically.

\begin{figure}[h]
\caption{A graph of the Linux snap installations (87 copies worldwide) shows the distribution of GNU Aris since February 2022 among various Linux based operating systems. Most users use Ubuntu 22.04 for the underlying operating system (this is shown in the middle region with the greatest area). Other users use mostly other Ubuntu systems or Fedora, Debian or Linux Mint.}
\label{fig:snap1}
\centering
\includegraphics[width=1\textwidth]{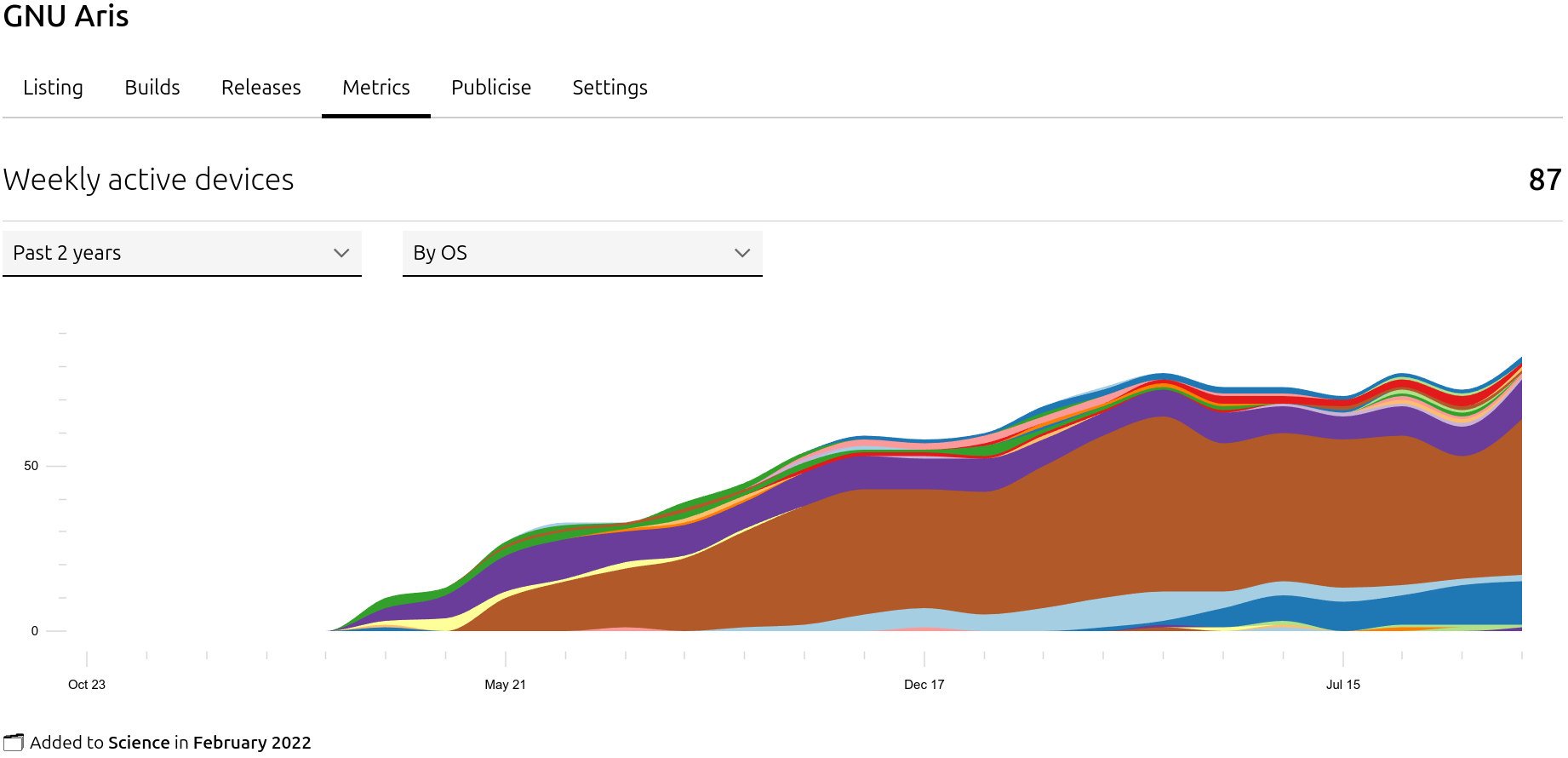}
\end{figure}

\begin{figure}[h]
\caption{Linux snap installations in October 2023 in a territorial distribution in 33 countries. For example, the territory ``United States'' has 19 installations.}
\label{fig:snap2}
\centering
\includegraphics[width=1\textwidth]{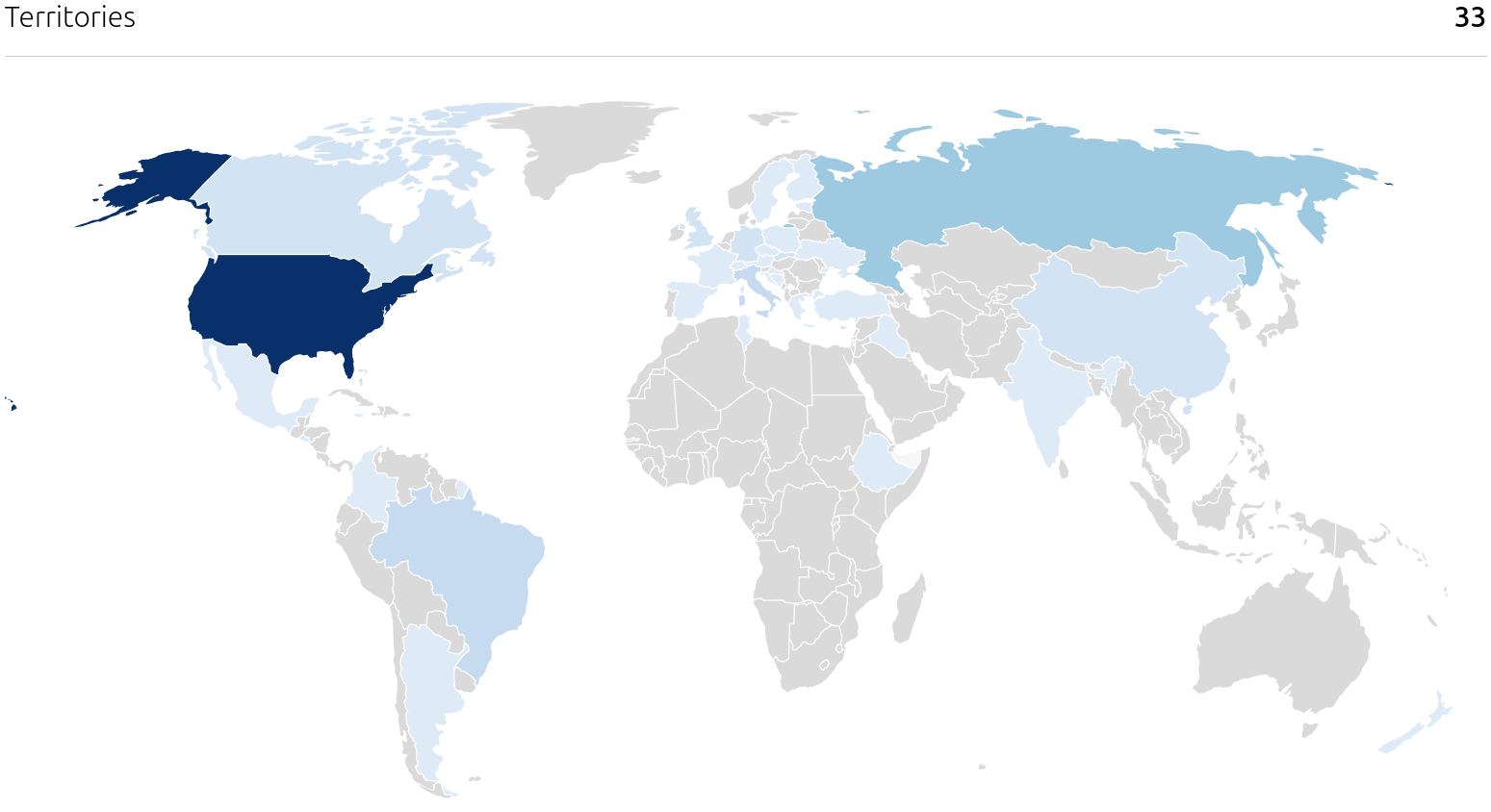}
\end{figure}

GNU Aris has been used during undergraduate lectures at PHDL, by prospective mathematics teachers,%
\footnote{\url{https://github.com/kovzol/logik/releases/download/April_2024/Logik.pdf}}
after it was
forked\footnote{\url{https://github.com/kovzol/aris}} and made available as a downloadable Windows
package\footnote{\url{https://github.com/kovzol/aris/releases/tag/2.2-22112022}} on GitHub. Its wide application among students was, however, difficult because the software required installation on either a Windows (\texttt{.exe} file) or Linux system
(snap,\footnote{\url{https://snapcraft.io}} see Figures \ref{fig:snap1} and \ref{fig:snap2}). In Austria, many students own a Mac computer. It was, therefore, very complicated for many students to install Aris since it usually meant that it had to be compiled from the sources, and such technical challenges cannot, usually, be addressed by non-technical users. Additionally, some Linux users also had to compile from the sources, since snaps are not enabled by default for many Linux distributions. 

Therefore, the second author, also a GNU activist, suggested defining a possible project under the umbrella of GNU, in the frame of the Google Summer of Code (GSoC) 2023
programme,\footnote{\url{https://summerofcode.withgoogle.com}} where a project supported by GNU had a good chance to participate. The first author of this paper applied to the programme and Google confirmed his participance. As a result, the first author worked 3 months very actively, to achieve the main goal: to make GNU Aris available for almost anyone, by porting the application to the web. His work was mentored by the second author and two other co-mentors, Andreas Ebetshuber and Alexander Thaller, both are students of PHDL.

\section{Technical description}

\begin{figure}[h]
\caption{GNU Aris checks the correctness of statements 7--13, based on premises 1--6.}
\label{fig:Aris alte Version}
\centering
\includegraphics[width=0.6\textwidth]{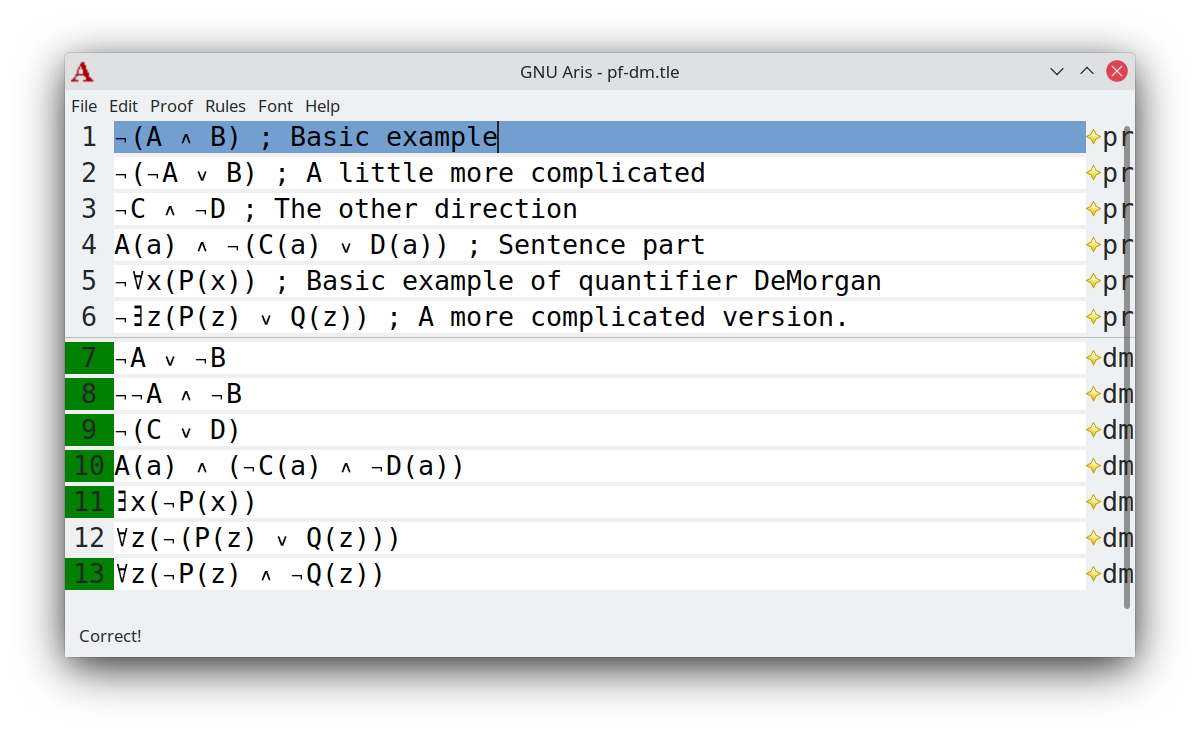}
\end{figure}

Therefore, the GSoC project aimed at extending GNU Aris by adding a web user interface. It involved some design changes in the graphical user interface (GUI), see Figure \ref{fig:Aris alte Version}, and migrating the old GTK-based user interface (UI) to Qt, a freely available framework provided by The Qt Company. Consequently, a native desktop along with a similar/ cross compiled web application based on HTML, Javascript, and WebAssembly was developed. It is also worth noting that it is possible to cross compile for Android, although it has not been done because the UI was not designed with mobile devices in mind. The separation of UI and logic enables interested programmers to implement their own UI if for example they wish to port Aris for Android.

GNU Aris had an old user interface that was based on GTK and did not support every platform. Since most of the source code was written around ten years ago, there also existed some amount of deprecated code, most of which dealt with GTK.

Before the official commencing of Google Summer of Code, several changes were made, which warranted another GTK version release. A bug fix disallowed references to the sub-proof from other lines except for the line immediately after the end of the sub-proof. The keyboard was also improved to allow keyboard input for Goals Window. We also changed the alignment of the three windows, i.e., Main Window, Rules Menu, and Keyboard, to look more symmetric on startup and synchronized window closing. For example, previously, the Rules and Keyboard had to be closed separately from the main window, which could cause them to stay open even after the main window was closed; this was fixed. After these changes, work began on the Qt version.

Redesigning a part of the UI and migrating to Qt could resolve some of the issues above and other GTK related problems. Also, since Qt has excellent support for generating WebAssembly code
(via the Emscripten\footnote{\url{https://emscripten.org}} toolchain), it seemed that a web application could be made more or less automatically after this major change.

When starting the GSoC project, it turned out that GNU Aris was forked by another team of developers at \url{https://github.com/Bram-Hub/aris}. It was first rewritten in Java and then in Rust. The Rust version can also be compiled into a web application, which can be tried out at \url{https://aris.bram-hub.com}, however it seems to be missing some features from the original version, like ``Import Proof'',
``Goals'', ``Export to Latex'' etc. Our port, however, still uses the original codebase written in C and adds some extensions in C++ \cite{cplusplus}.

\section{Results}

Between May and August 2023, the first author managed to address the above-mentioned challenges. His technical summary can be found at \url{https://github.com/kovzol/aris/pull/15}. Here some of his achievements are summarized.

The following goals have been met:
\begin{itemize}
\item Design Changes
\item Build Qt-based Application
\item Build WebAssembly Application
\item Dark Mode for Qt and WebAssembly versions
\end{itemize}

We made most of the design decisions during the Community Bonding Period (May 2023).

For the first half of the project, the biggest work was performed on the GUI with the Qt Modeling Language (QML\footnote{\url{https://doc.qt.io/qt-6/qtqml-index.html}}). QML is a user interface markup declarative language for designing user interface-centric applications.

Then, in July onwards we worked on connecting the GUI with the C logic engine. We decided to leverage the Qt Quick C++ API  to add an intermediary, written in C++, to handle communications between the UI and the Logic Engine. The QML-C++ side of the connection proceeded quite smoothly, thanks to the robust Qt Quick framework. The C-C++ side was more complicated where we used pre-existing C source files wherever possible, by declaring linkage specifications using preprocessor macros to create mixed-language header files which can be compiled with both C as well as compatible C++ compilers and linked together. However, in some cases, the first author had to write the C implementation logic in the C++ layer himself, because the source files had it bundled with the GTK implementation. This decision can be useful for future implementations (based on platforms other than Qt) since now we have a mostly GUI-independent logic engine written purely in C.

Thus, the architecture looks somewhat like this (see also Figure \ref{fig:architecture}):

\begin{figure}[h]
\caption{An architecture graph of the data flow}
\label{fig:architecture}
\centering
\includegraphics[width=1\textwidth]{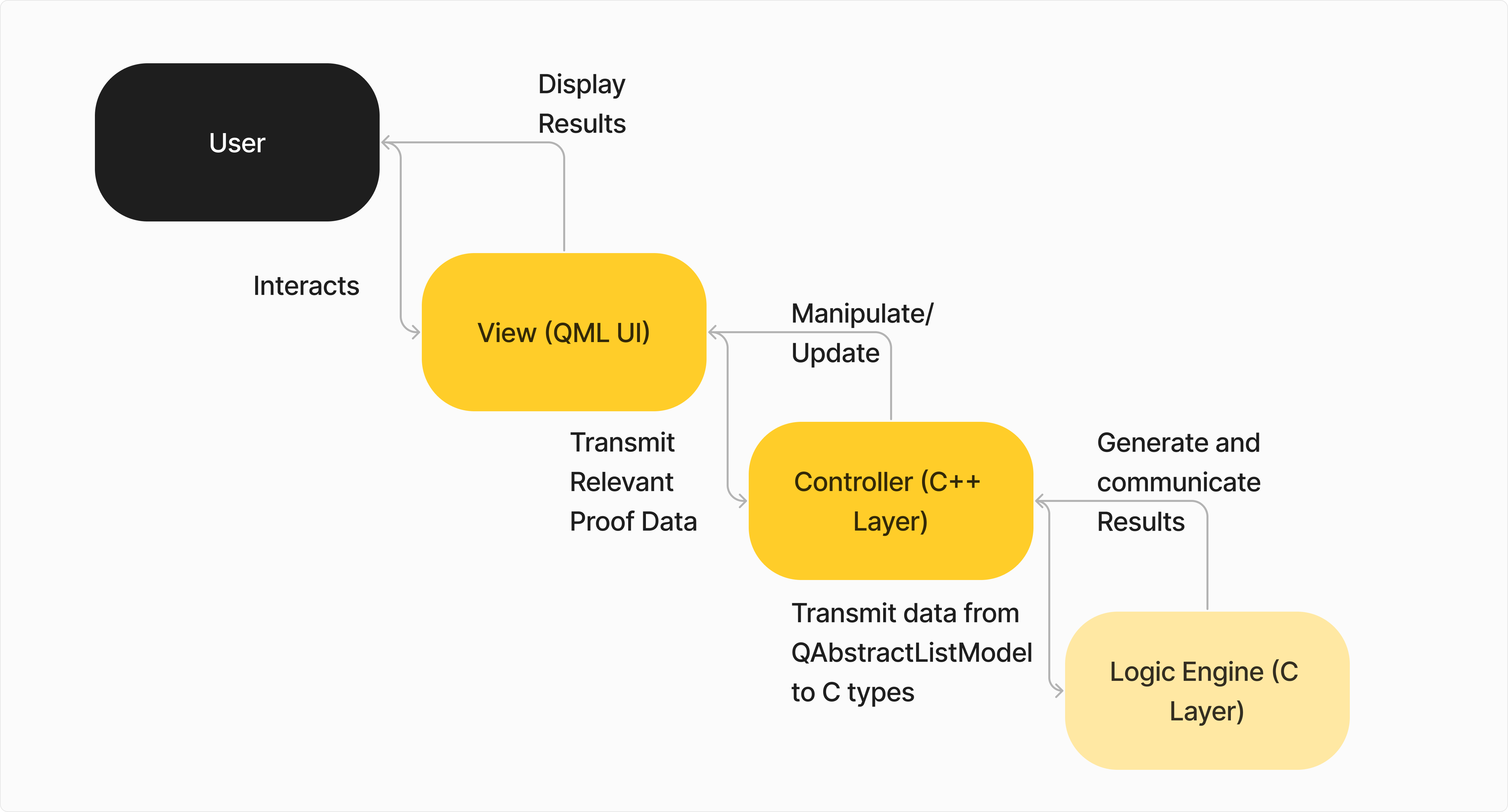}
\end{figure}

\begin{itemize}
\item a QML-based top-level GUI that communicates key events and data with the C++ layer,
\item a C++ middle layer that stores data in QAbstractListModel and communicates with both QML and C logic layer,
\item a C logic layer that receives proof data and generates results that are communicated back to the C++ layer.
\end{itemize}

Finally, the compilation to WebAssembly (via Emscripten) was relatively smooth. We used the fusion style for consistency across Desktop and WebAssembly builds.

Also, since the WebAssembly (wasm) code is run in a sandboxed environment, native QML FileDialogs were useless which caused problems with downloading and uploading proofs for the web. Thankfully, it was possible to implement Qt's \texttt{QFile\-Dialog::\-get\-Open\-File\-Content} API to get around the problem. As a result, most options (like \textit{Open} and \textit{Import Proof}) have a separate wasm function in the middle C++ layer (that uses the same native function underneath after resolving file I/O). We could have overloaded the same function, but this is more readable.

We also faced a couple of problems. During compilation to WebAssembly, we needed clarification about many of the errors thrown. So, we learned, for example, about compiling libraries and how dependencies work. We had to compile \texttt{libxml2} from source with Emscripten because it was not originally provided. Our efforts in this direction could be helpful for other developers looking to compile \texttt{libxml2} with the Emscripten toolchain. Thus, finally, in August, we had a working Aris website hosted by the second author.

\begin{figure}[h]
\caption{Web version of GNU Aris checks the correctness of statements 5--8 based on premises 1--4.}
\label{fig:Aris Web-Version}
\centering
\includegraphics[width=1\textwidth]{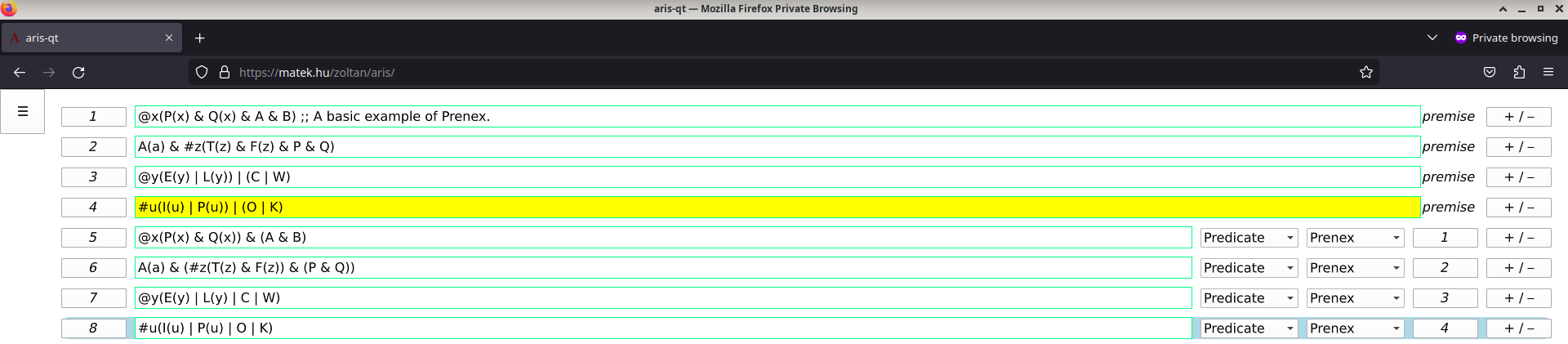}
\end{figure}

Figure \ref{fig:Aris Web-Version} shows how the WebAssembly implementation is performing in Mozilla Firefox after it is deployed at its current official web page \url{https://matek.hu/zoltan/aris}. 

We have also managed to create a Linux AppImage\footnote{\url{https://appimage.org}} for the Qt native version which will allow users to directly run Aris natively without compiling. A Windows executable is also in progress.

\begin{figure}[h]
\caption{The native Qt version of GNU Aris checks the correctness of statements 7--13, based on premises 1--6.}
\label{fig:Aris neue Version}
\centering
\includegraphics[width=1\textwidth]{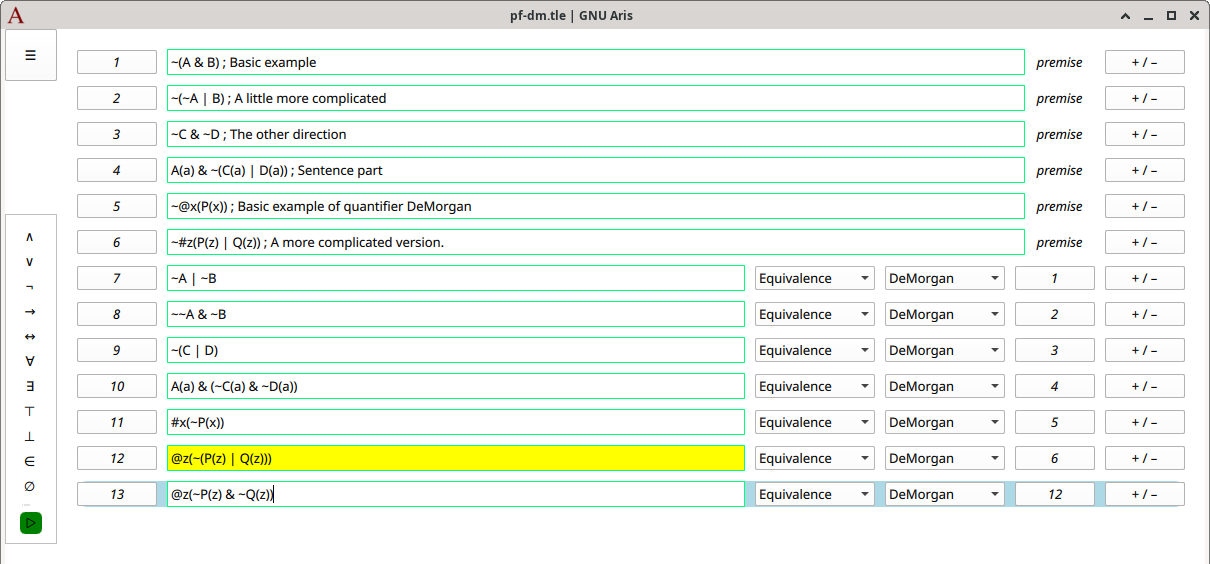}
\end{figure}

Also, Figure \ref{fig:Aris neue Version} shows the same example as in Figure \ref{fig:Aris alte Version}, but in the native version of the Qt UI.

\section{The role of GNU Aris in education}

The education of mathematics teachers at primary level and secondary level includes an understanding of logical deduction and mathematical proofs. Even though the PHDL and cooperating institutions for the education of teachers at primary level do not share a common curriculum, each curriculum provides a detailed description for the mathematical education of prospective teachers, that mentions subject-specific and educational
skills,\footnote{\url{https://www.phdl.at/fileadmin/user_upload/3_Service/2_Studienbetrieb/Mitteilungsblatt/MB-012-2018-20180621_CurrPrimar_BAC_MA_PH_Linz-veroeffentlicht.pdf}} the importance of logic and mathematical
reasoning\footnote{\url{https://qmpilot.phsalzburg.at/File/CoreDownload?id=2170&filename=PG_098_099_Curriculum_Primarstufe_BA_MA-60_MA-90_red.pdf&langId=1}}
as well as the knowledge and understanding of theorems in geometry and
arithmetic.\footnote{\url{https://www.kph-es.at/fileadmin/user_upload/Curriculum_Bachelor-_und_Masterstudium_Lehramt_Primarstufe2022.pdf}} Furthermore, all cooperating institutions share a common curriculum for the education of mathematics teachers at secondary level, in which logical deduction and the knowledge of commonly used methods for mathematical proofs are considered basic knowledge and prerequisites for courses on analysis, algebra, geometry, statistics and
stochastics.\footnote{\url{https://www.liles.at/fileadmin/user_upload/BEd_2021.pdf}}

Aris may be utilized to improve the understanding of logical deduction and mathematical proofs, including premises and necessary steps, that lead to correct conclusions. The program requires the user to actively work with propositional and predicate logic, Boolean algebra and arithmetical logic combined with commonly used methods of mathematical proofs, like proof by contradiction and proof by mathematical induction, while also following rules of inference (e.g.~modus ponens).

The program is still in development, however, direct comparison with other published state of the art proof assistants shows, that there are some differences, which could be vital for the implementation of Aris in education of mathematics teachers at primary level and secondary level. 

Following professional proof assistant programs were compared with Aris: 
\begin{itemize}
    \item Isabelle\footnote{\url{https://isabelle.in.tum.de/}} (Technical University of Munich, University of Cambridge)
    \item Coq\footnote{\url{https://coq.inria.fr/}} (Institut national de recherche en informatique et en automatique)
    \item Lean\footnote{\url{https://lean-lang.org/}} (Lean Focused Research Organization)
    \item HOL Light theorem prover\footnote{\url{https://www.cl.cam.ac.uk/~jrh13/hol-light/}} (John Harrison)
    \item The Incredible Proof Machine\footnote{\url{http://incredible.pm/}} (Joachim Breitner)
    \item QED -- an interactive textbook\footnote{\url{https://teorth.github.io/QED/}} (Terence Tao)
\end{itemize}

These programs may be utilized to support proving mathematical theorems formally in higher-order logic and some of them are even available in form of web applications. Nevertheless, proof assistants like Isabelle, Coq, Lean and HOL Light theorem prover require the user to have basic programming skills for inputs with correct syntax. Furthermore, users have to work with designated metalanguages depending on the program used. Since the curricula for the education of mathematics teachers of primary level and secondary level do not comprise the acquisition of basic programming skills besides of elective courses, students may lack necessary knowledge to be able to use these programs. As a result, the implementation of proof assistants based on metalanguages and syntax in education of prospective mathematics teachers would take up time to learn necessary skills. Aris provides a rather intuitive user interface for both propositional and predicate logic, which leads to time efficient application of the program for education. Other programs like The Incredible Proof Machine and QED -- an interactive textbook enable users to improve their knowledge of logical deduction and mathematical proofs in rather playful environments without further knowledge of syntax for programming-based proof assistants. Such programs provide easy-to-use interfaces in web applications, which could be suitable for means of education. These proof assistants offer predefined exercises in modular course systems, which are advantageous for the introduction of students to propositional and predicate logic. However, the access to a limited number of different exercises is a restriction for the improvement of skills in logical deduction and mathematical proofs. For the purpose of further honing those skills, Aris may be the preferred proof assistant program, since its utilization is not limited to certain exercises.

A group of prospective mathematics teachers at the PHDL tried out the web application of GNU Aris during a course on mathematical logic in the winter semester of 2023/24. Some examples were given to explain how to work with the program. The introduction included the explanation on correct statement inputs as premises, adding conclusions and sub-proofs with their designated assumptions and how to prove each added conclusion by selecting specific rules of inference or correct equivalences.
Then, a chapter of Smullyan's \textit{The Lady or the Tiger} was presented to the group of students, where prisoners are put on trial in form of logical problems. These trials involve multiple rooms, which are labelled with statements about these rooms. Furthermore, the prisoners are told facts about the rooms and additional rules by the king, who is putting them on trial. The prisoners must then proceed to pick one of the rooms, knowing, that there might be either a lady or a tiger behind the door. In case of picking a room with a lady inside, the prisoner is set free and allowed to marry that lady, but opening the door of a room with a tiger inside might lead to the prisoner being eaten \cite{tlott}. In total there are twelve different trials, which can be solved by logical thinking. One of the first and easier trials was chosen to show the students, how to use Aris to formalize the given problem and how to resolve that problem while proving each step that leads to the correct conclusion. After the example, the students were supposed to choose and solve one of the other trials described in the book individually. The goal of that exercise was getting used to the proof assistant program while correctly formalizing and solving a logical puzzle.

\begin{figure}[h]
\caption{The German translation of the fifth trial, which was chosen by one of the students.}
\label{fig:ger5}
\centering
\includegraphics[width=0.6\textwidth]{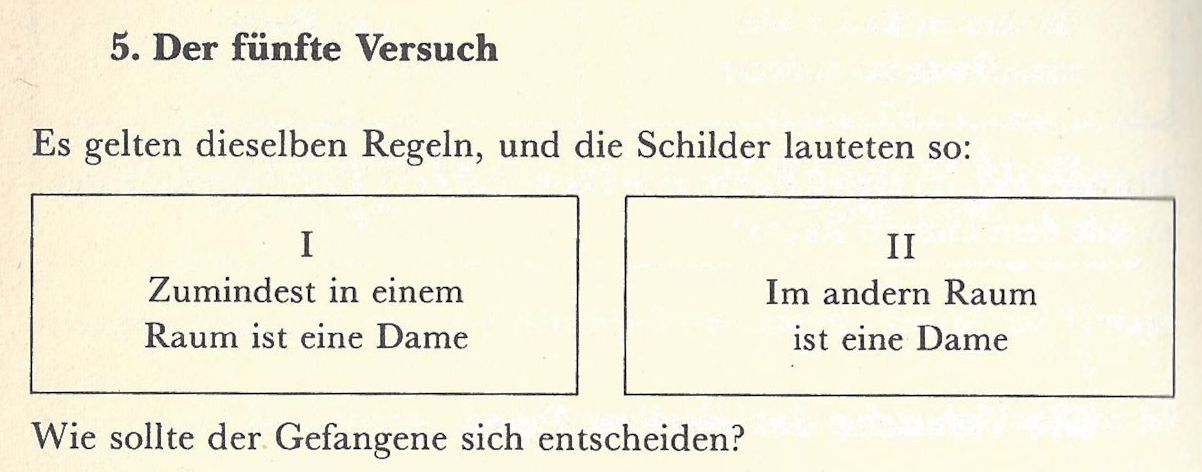}
\end{figure}

One of the students decided to submit a solution for the fifth trial mentioned in the book, where a decision between two different rooms had to be made. The first room is labeled with a sign stating that at least one of the two rooms contains a lady, while the sign on the second door states that the first room contains a lady. Additionally, the king applies a rule, which stipulates that a lady in the first room means the sign of that room is true, while a tiger in that room means the sign false. The opposite of that rule is applied to the second room, so that a tiger in the second room means the sign of that door is true and a lady inside means the sign is false \cite{tlott}. 
Figure \ref{fig:ger5} shows the original problem shown to the group of students, which originates from the German translation of Smullyan's \textit{The Lady or the Tiger} \cite{dot}.

\begin{figure}[h]
\caption{The student's premises and the sub-proof leading to the contradiction of a lady being in the second room.}
\label{fig:subp5}
\centering
\includegraphics[width=1\textwidth]{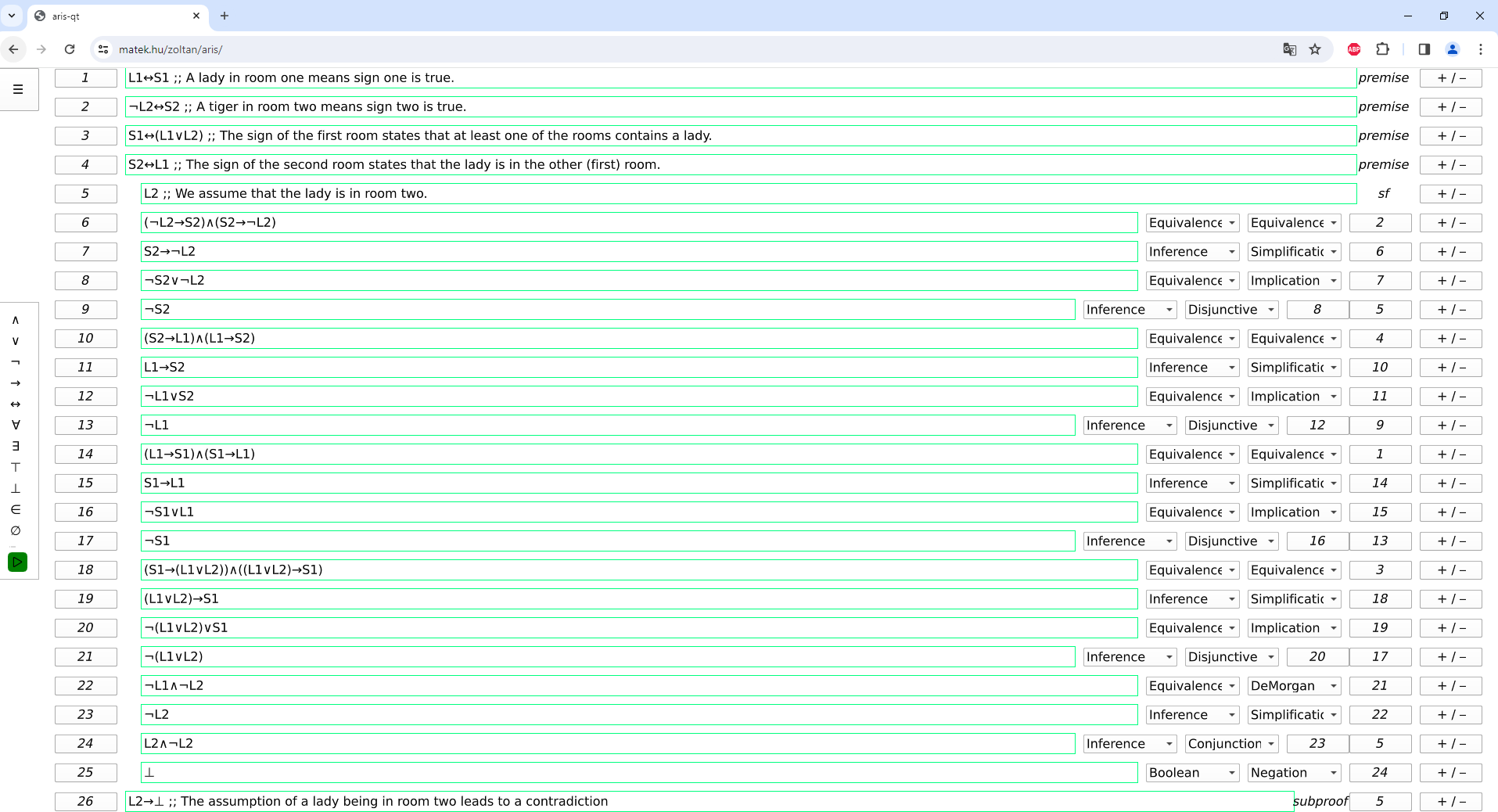}
\end{figure}

The student formalized the statements of each sign and the rules as premises with Aris. For that purpose, all statements were numbered according to the room they belong to. Furthermore, statements for the signs being true (\(S\)), the negation (\(\neg S\)), pointing out a false sign, a lady in the designated room (\(L\)) and the negation (\(\neg L\)), pointing out the presence of a tiger in the room, were used. The student then proceeded with a sub-proof to show that the assumption of a lady being in the second room leads to a contradiction. Figure \ref{fig:subp5} shows the student's approach of using the web application of Aris for said sub-proof and concluding, that there has to be a tiger in the second room.

\begin{figure}[h]
\caption{The student's proof that a lady is in the first room and a tiger is in the second room.}
\label{fig:proof5}
\centering
\includegraphics[width=1\textwidth]{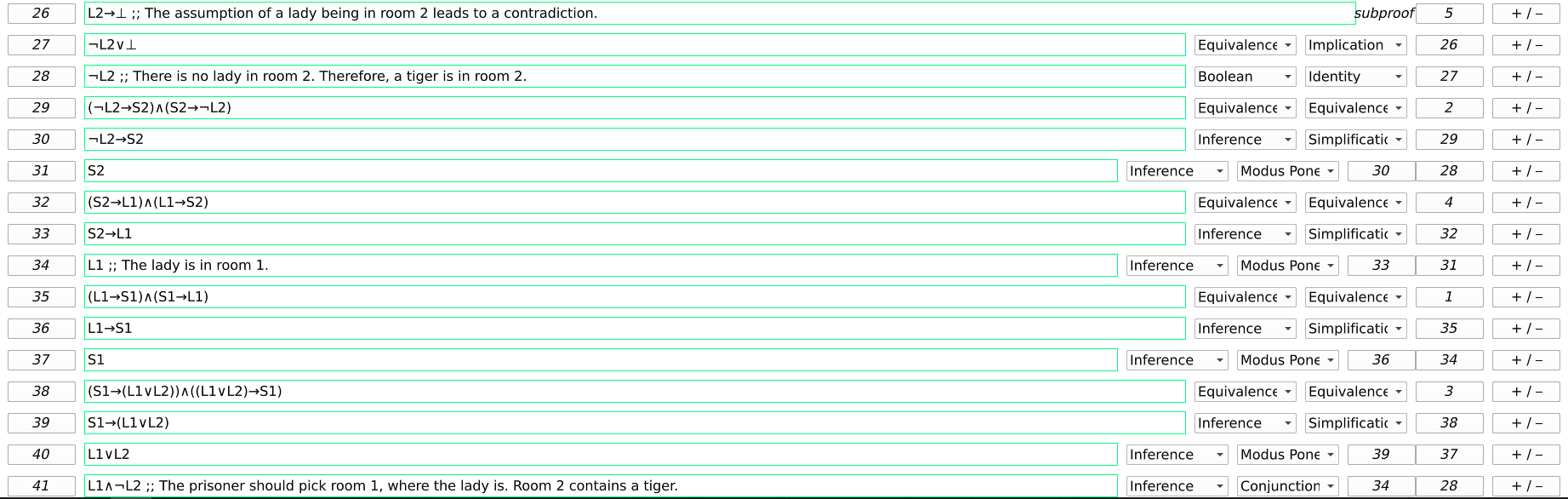}
\end{figure}

The student concluded, that a tiger in the second room means that the sign attached to the door must be true and continued proving that there is a lady in the first room, as shown in Figure \ref{fig:proof5}. The correct example solution of the student consisting of the sub-proof and the following conclusion shows, that rules of inference, equivalences of propositions and Boolean algebra were needed to prove every step leading to the solution of the logical problem.

The procedure of introducing GNU Aris to prospective mathematics teachers by resolving humorous logical problems was very successful, since many students presented a variety of correct solutions made via the web application. The group of prospective mathematics teachers managed to deliver great results while gaining experience with the program to be built upon for further utilization of Aris. Nevertheless, some students did not manage to complete their task of solving one of the trials. Student's reports included technical issues while working with proof assistant, while other students required additional instructions on how to use Aris. Especially checks of the correctness of certain statements led to error messages, which were not sufficient for the students to find out the cause of the error. Therefore, all students of the group were invited to deliver feedback for future improvement of GNU Aris to enhance the usability and accessibility of the proof assistant program. 

Some enthusiastic students continued working with the program to contribute to the improvement of the experience for future students by providing feedback for technical issues and suggestions for quality enhancements of the program. GNU Aris does have the feature to save inputs made, but it has repeatedly been reported that the correct file extension is not added automatically during that process, which renders saved files obsolete and therefore leads to the loss of work by the students who did not notice the missing file extension. Another technical issue was reported, when users had already input multiple lines and were required to scroll up to select other statements. In some cases, the program snaps back to the starting point of the scrolling. 

Further submissions by the group of students suggest following possibilities for refinements of GNU Aris for future users:
\begin{itemize}
    \item more comprehensible and detailed error messages,
    \item automatic updates on the feedback for the correctness of statements after changing faulty inputs,
    \item a feature to zoom for smaller device displays,
    \item an option to change the user language,
    \item an option to highlight, copy and paste inputs,
    \item an option to change the type of input between premise and conclusion after inserting a new line,
    \item an operator for exclusive disjunction,
    \item an option to use the rule of inference for a proof by contrapositive.
\end{itemize}

\begin{figure}[h]
\caption{The suggested solution for the plenary exercise.}
\label{fig:jimmy}
\centering
\includegraphics[width=1\textwidth]{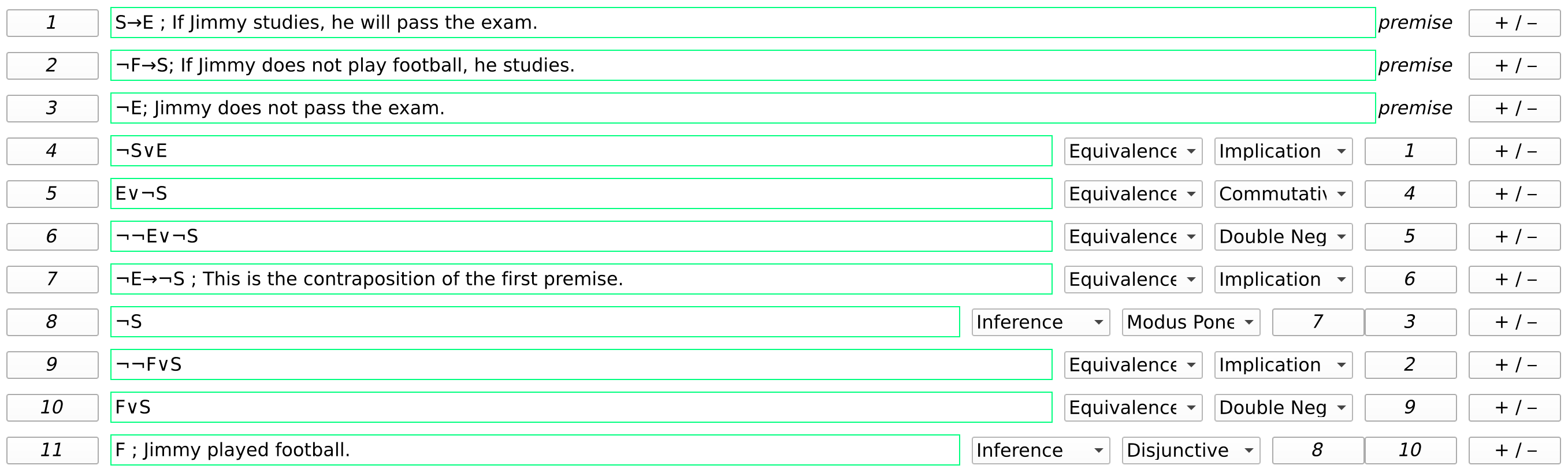}
\end{figure}

Recently, GNU Aris was introduced to a different group of 15 students of teaching during another course at the PHDL in the winter semester of 2024/25. After a brief explanation on how to use Aris, the students were instructed to use the given premises to make deductions in a plenary exercise. Figure \ref{fig:jimmy} shows a translation of the exercise with commented inputs suggested by the students (with some hints given by the lecturer), using equivalences and rules of inference, finally leading to a correct conclusion. This exercise illustrates that GNU Aris can be a helpful tool for teaching logical deduction, as students were able to complete the task with minimal guidance.

It is intended that GNU Aris is straightforward to use, requiring only basic knowledge of logical deduction and mathematical proofs. This makes it potentially suitable for use in the education of mathematics teachers at the primary and secondary levels. While most students were able to work with the program successfully, some encountered difficulties and needed additional support, indicating areas where the tool could be improved for future use.

\section{Future plans and conclusion}

\begin{figure}[h]
\caption{\LaTeX~export of a valid proof in first-order logic.}
\label{fig:latex}
\centering
\includegraphics[width=0.5\textwidth]{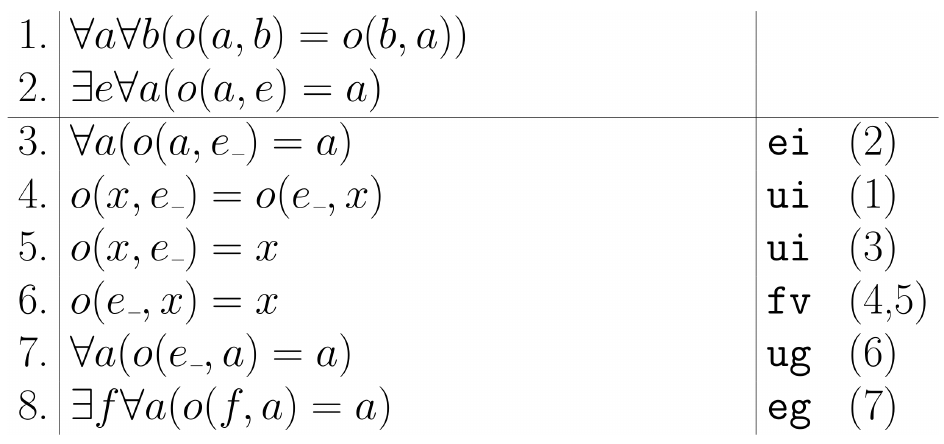}
\end{figure}

At the PHDL, a short overview on first-order logic was given by using the program: Given a commutative binary operation $o$ and its right identity $e$, one needs to prove that there exists a left identity $f$ as well. Figure \ref{fig:latex} shows how a valid proof can be exported as a \LaTeX~file and later inserted into a mathematical document. First-order logic derivation was shown, however, just as a test-case and was not explained in further detail. There are plans to further integrate GNU Aris into future iterations of the course on mathematical logic for prospective mathematics teachers, with anticipated course improvements eventually including a more comprehensive introduction to first-order logic using GNU Aris. By incorporating GNU Aris, the course aims to provide students with hands-on experience in constructing and validating logical proofs, helping to solidify their foundational understanding of formal reasoning processes. In addition, the program will be used as a tool to assess whether it effectively supports students' comprehension and application of logical reasoning principles.

Although the original goal, that is, to port GNU Aris to be a full-featured standalone web application to increase the accessibility of users with different devices and operating systems, has already been met, we would like to continue improvements of Aris in the following points:

\begin{enumerate}
\item Update icons (to improve visibility in dark mode)
\item Make a short how-to-use tutorial
\item Add a SAT Solver \cite{sat}
\item Finish Isar\footnote{\url{https://isabelle.in.tum.de/doc/isar-ref.pdf}} interoperability
\end{enumerate}

Additionally, received feedback from the group of students will be considered for additional enhancements to tailor the program and its interface specific to the purpose of being used for the education of mathematics teachers of primary level and secondary level.
When the new GUI of GNU Aris is tested again in the future by groups of prospective mathematics teachers at the PHDL, they will also be asked for feedback to further improve the proof assistant program.

\section{Acknowledgments}

Second author collaborated with some participants of
the grant ``Augmented Intelligence in
Mathematics Education through modeling, automated reasoning and
artificial intelligence'' (IAxEM-CM/PHS-2024/PH-HUM-383), supported by
Comunidad de Madrid.

We are grateful to the GNU Project and Google by supporting our work during the GSoC 2023 project. Special thanks to Jose E.~Marchesi, GNU's organization admin, for his kind encouragement. Also, we are grateful to co-mentors Andreas Ebetshuber and Alexander Thaller for their help during the project.

\nocite{*}
\bibliographystyle{eptcs}
\bibliography{generic}
\end{document}